\documentclass[twocolumn,prb,showpacs,superscriptaddress,amsmath,amssymb]{revtex4-1}
\usepackage{graphicx}
\usepackage{dcolumn}
\usepackage{bm}
\usepackage{amsmath}
\usepackage{amssymb}
\usepackage[usenames]{color}
\usepackage[normalem]{ulem}
\usepackage[none]{hyphenat}
\usepackage{soul}
\usepackage{float}

\begin{document}


\title{Competing rhombohedral and monoclinic crystal structures in Mn$Pn_2Ch_4$ compounds: an {\em ab-initio} study}

\author{S.~V. Eremeev}
 \affiliation{Institute of Strength Physics and Materials Science,
634021, Tomsk, Russia}
 \affiliation{Tomsk State University, 634050 Tomsk, Russia}
 \affiliation{Saint Petersburg State University, Saint Petersburg, 198504,
 Russia}

\author{M.~M. Otrokov}
\affiliation{Donostia International Physics Center (DIPC),
             20018 San Sebasti\'an/Donostia, Basque Country,
             Spain}
 \affiliation{Tomsk State University, 634050 Tomsk, Russia}
 \affiliation{Saint Petersburg State University, Saint Petersburg, 198504,
 Russia}

\author{E.~V. Chulkov}
\affiliation{Donostia International Physics Center (DIPC),
             20018 San Sebasti\'an/Donostia, Basque Country,
             Spain}
\affiliation{Departamento de F\'{\i}sica de Materiales UPV/EHU,
Centro de F\'{\i}sica de Materiales CFM - MPC and Centro Mixto
CSIC-UPV/EHU, 20080 San Sebasti\'an/Donostia, Basque Country, Spain}
 \affiliation{Tomsk State University, 634050 Tomsk, Russia}
 \affiliation{Saint Petersburg State University, Saint Petersburg, 198504,
 Russia}

\begin{abstract}
Based on the relativistic spin-polarized density functional theory
calculations we investigate the crystal structure, electronic and
magnetic properties of a family Mn$Pn_2Ch_4$ compounds, where
pnictogen metal atoms ($Pn$) are Sb and Bi; chalcogens ($Ch$) are
Se, Te. We show that in the series the compounds of this family with
heavier elements prefer to adopt rhombohedral crystal structure
composed of weakly bonded septuple monoatomic layers while those
with lighter elements tend to be in the monoclinic structure.
Irrespective of the crystal structure all compounds of the
Mn$Pn_2Ch_4$ series demonstrate a weak energy gain (of a few meV per
formula unit or even smaller than meV) for antiferromagnetic (AFM)
coupling for magnetic moments on Mn atoms with respect to their
ferromagnetic (FM) state. For rhombohedral structures the interlayer
AFM coupling is preferable while in monoclinic phases intralayer AFM
configuration with ferromagnetic ordering along the Mn chain and
antiferromagnetic ordering between the chains has a minimum energy.
Over the series the monoclinic compounds are characterized by
substantially wider bandgap than compounds with rhombohedral
structure.
\end{abstract}

\maketitle

\section{Introduction}


The ternary chalcogenides $MPn_2Ch_4$ ($M$ = Fe, Mn; $Pn$ = Sb, Bi;
and $Ch$ = S, Se)
\cite{Bente,Buerger,Djieutedjeu,Leone,Matar,Tian,Wintenberger,Ranmohotti,Djieutedjeu_EJICh2011}
which include a large number of synthetic and natural metal
chalcogenides are promising for applications in thermoelectricity
 \cite{Rowe}, spintronics \cite{Wolf}, and nonlinear
optics \cite{Ballman}. These compounds crystallize in the monoclinic
space group $C2/m$. This structure is characterized by a presence of
the $M$ atoms chains along one of the crystallographic directions
where the distances between $M$ atoms are 1.5-2 times shorter than
in other directions. As shown by magnetic measurements, the iron
containing chalcogenides are ferromagnetic semiconductors
\cite{Djieutedjeu}, while the Mn-based compounds are
antiferromagnets \cite{Djieutedjeu_EJICh2011}. However, recently a
new ternary chalcogenide semiconductor of the same series containing
a heavier chalcogen atom (Te), MnBi$_2$Te$_4$, and possessing
different crystal structure has been reported \cite{DSLee}. It was
shown that the compound crystallizes in the rhombohedral structure
($R\bar3m$) and can be described as the one composed of septuple
layer (SL) slabs with a stacking sequence of
Te1--Bi--Te2--Mn--Te2--Bi--Te1 along the $c$-axis and with van der
Waals gaps between the slabs. The structure may also be described by
using the Bi$_2$Te$_3$ structure, where the central Te layer is
substituted with Te--Mn--Te layers. The obtained phase, as
established by using high-temperature XRD analysis, is stable up to
423 K while above this temperature it starts to be spontaneously
decomposed into Bi$_2$Te$_3$ and MnTe$_2$ phases. It should be noted
that the magnetic state of MnBi$_2$Te$_4$ has not been studied.

In the present study we scrutinize the crystal structure of
compounds of the Mn$Pn_2Ch_4$ series ($Pn$ = Sb, Bi; $Ch$ = Se, Te)
by means of the density functional theory (DFT) calculations with
taking into account their magnetic state.

\begin{figure*}
\includegraphics[width=\textwidth]{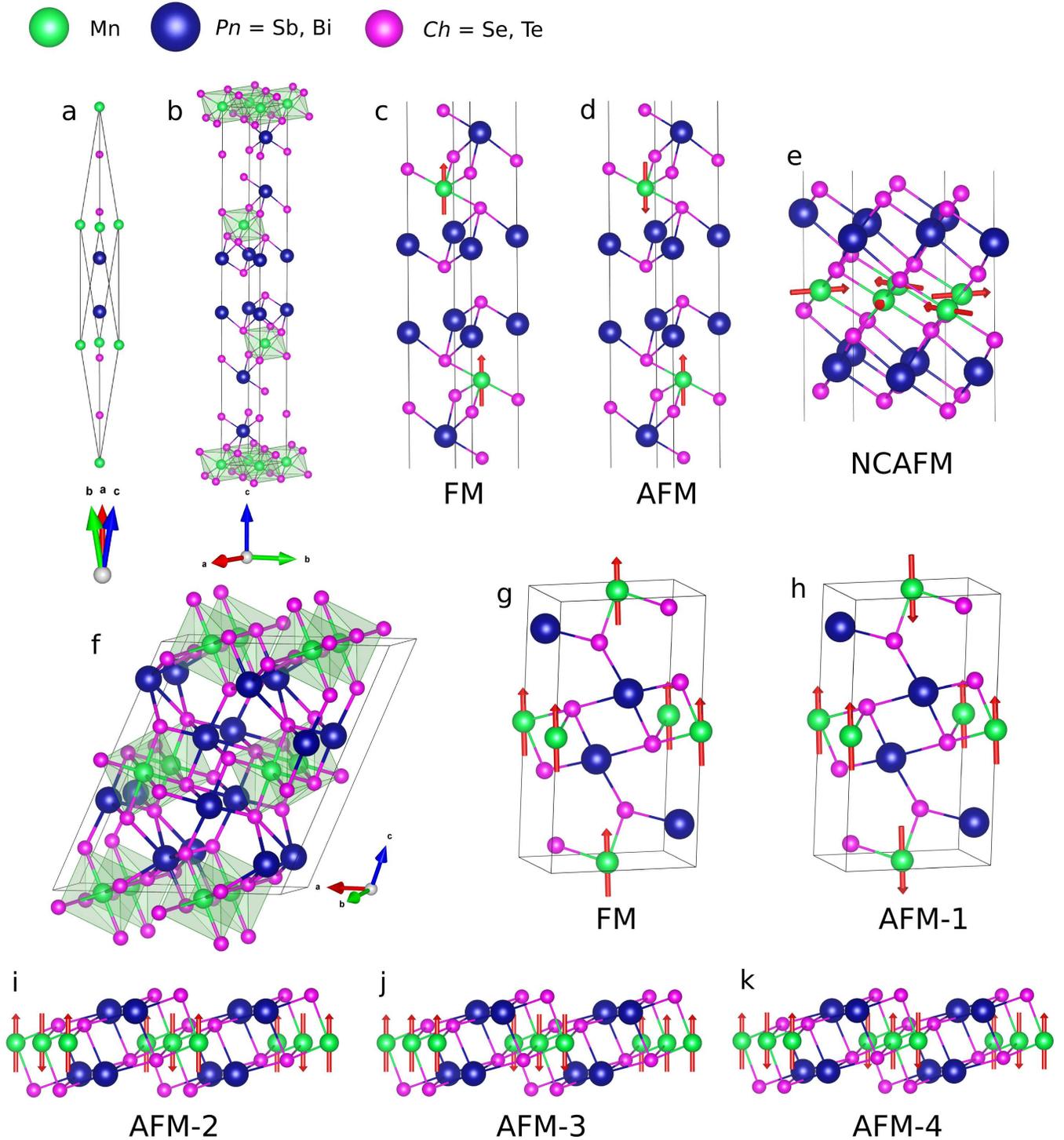}
 \caption{Rhombohedral (a) and respective hexagonal (b)
representations of the crystal structure as was found for
MnBi$_2$Te$_4$ compound in Ref.~\citenum{DSLee}. Parts of the
doubled along $c$ axis hexagonal cells, containing two SL blocks
with FM (c) and AFM (d) ordering for Mn atoms (red arrows show
mutual orientations of the magnetic moment). (e) NCAFM ordering in
the $\sqrt 3$$\times$$\sqrt 3$ hexagonal cell. (f) Monoclinic cell
as was found for MnBi$_2$Te$_4$ compound in
Ref.~\citenum{Ranmohotti}. Triclinic Niggli-reduced cell for the
monoclinic structure with FM (g) and interlayer AFM (AFM-1) (h)
magnetic ordering. (i-k) Intralayer magnetic configurations:
antiferromagnetically ordered Mn chains with ferromagnetic
interchain coupling (AFM-2), ferromagnetically ordered chains with
antiferromagnetic interchain coupling (AFM-3), and checkerboard-like
magnetic configuration with antiferromagnetic both intrachain and
interchain coupling (AFM-4).}
 \label{struct}
\end{figure*}

\section{Methods}

For calculations we use the Vienna Ab Initio Simulation Package
(VASP) \cite{VASP1,VASP2} with generalized gradient approximation
(GGA) \cite{PBE} for the exchange correlation potential. The
interaction between the ion cores and valence electrons is described
by the projector augmented-wave method \cite{PAW1,PAW2}. Owing to
presence of heavy $Pn$ atoms (Sb,Bi) in the ternary chalcogenides
$MPn_2Ch_4$ the relativistic effects are expected to have a
significant impact on electronic and crystal structures. For this
reason the relativistic effects, including spin-orbit interaction,
were taken into account in the calculations. To determine
equilibrium state of the systems, we accurately optimized the
lattice parameters as well as the atomic positions of the
rhombohedral and monoclinic structures of Mn$Pn_2Ch_4$ in
ferromagnetic (FM) and different antiferromagnetic states. For
rhombohedral structure we have considered a collinear
antiferromagnetic state with interlayer antiferromagnetic coupling
(AFM) and noncollinear antiferromagnetic (NCAFM) state in which
three spin sublattices form angles of 120$^\circ$ with respect to
each other \cite{Lee.prb1986}. To consider the FM and AFM ordering
in the rhombohedral phase we use doubled along $c$ axis hexagonal
cell containing six Mn atoms. The NCAFM configuration was treated
within larger supercell, which is $\sqrt 3$$\times$$\sqrt 3$ in the
hexagonal plane. For monoclinic phase we studied four AFM
configurations on the base of triclinic Niggli-reduced cell,
containing two Mn atoms, which was 2$\times$2 expanded in the basal
plane. The considered AFM configurations are the interlayer
antiferromagnetic configuration with antiferromagnetic coupling
through the longest Mn-Mn distance (along monoclinic $c$ axis)
between ferromagnetic layers (AFM-1) and three intralayer
configurations: antiferromagnetically ordered Mn chains with
ferromagnetic interchain coupling (AFM-2), ferromagnetically ordered
chains with antiferromagnetic interchain coupling (AFM-3), and
checkerboard-like magnetic configuration where both intrachain and
interchain coupling are antiferromagnetic (AFM-4). DFT-D3 van der
Walls corrections \cite{Grimme} were applied for accurate structure
optimization. To describe the strongly correlated Mn-$d$ electrons
we include the correlation effects within the GGA+$U$ method in the
Dudarev implementation \cite{Dudarev}. Since the nearest neighbors
for Mn atom in both rhombohedral and monoclinic structures are
Te(Se) atoms we have chosen the $U^*=U-J=$5.34(5.33) eV  values to
be the same as in bulk MnTe(MnSe) \cite{Youn}.

\section{Results and discussion}

\subsection{MnBi$_2$Te$_4$}

The MnBi$_2$Te$_4$ compound was grown for the first time and its
crystal structure was determined in Ref.~\citenum{DSLee}. It was
found to crystallize in the rhombohedral crystal phase (space group
$R\bar3m$) and have a layered structure composed of septuple layer
(SL) slabs with a stacking sequence of
Te1--Bi--Te2--Mn--Te2--Bi--Te1 along the hexagonal axis with van der
Waals gaps between them (Fig.~\ref{struct} (a,b)). According to the
powder X-ray diffraction (XRD) data \cite{DSLee} the unit cell
parameters are $a$ = 4.334 \AA\ and $c$ = 40.910 \AA. The atomic
positional parameters were also determined. The DFT calculations
performed in Ref.~\citenum{DSLee} reproduced well the $a$ parameter
while noticeably overestimated (by 5.06~\%) the $c$ one. Such an
overestimation  is typical for the DFT calculations performed for
the  layered structures without taking the van der Waals corrections
into account. Besides, only FM state was considered in that
calculation. In order to find out the ground state of MnBi$_2$Te$_4$
we consider FM, AFM, and NCAFM magnetic ordering in the rhombohedral
structure (Figs.~\ref{struct} (c--e)) as well as FM and four above
described AFM alignments in the monoclinic structure
(Figs.~\ref{struct} (g--k)) which is typical for Mn$Pn_2Ch_4$ ($Ch$
= S, Se). For each configuration the lattice parameters and atomic
positions were optimized. According to the calculations the lowest
energy structure is the rhombohedral one with an AFM interlayer
coupling of the Mn magnetic moments which were found to be
4.607~$\mu_{\rm B}$. The obtained equilibrium lattice constants
$a$=4.336 \AA\ and $c$=40.221 \AA\ as well as the atomic positions
(Table~\ref{tab1}) agree well with the experimental parameters. It
is worth to note that other magnetic configurations of the
rhombohedral structure, FM and NCAFM, are 4.5 meV and 11.8 meV per
formula unit, respectively higher in energy than the AFM ground
state. On the other hand the FM (Fig.~\ref{struct} (g)) and
different AFM configurations (Figs.~\ref{struct} (h--k)) of the
monoclinic structure have the total energy of more than 200 meV
higher than that of the ground state although they differ between
themselves by few meV only. We remind, that all calculations were
done for the same $U^*$ value as in bulk MnTe (see Methods section).
Additionally, an extensive testing was performed for the
rhombohedral phase in order to ensure stability of the results
against the $U^{*}$ value change. At that, the crystal structure was
fully optimized for each $U^{*}$ considered. It was found that
neither intra- nor interlayer magnetic ordering changes
qualitatively when $U^{*}$ varies from 3 to 5.34 eV and the energy
gain for AFM phase as compared to ferromagnetic ordering is larger
for smaller $U^*$.

\begin{table}[t]
\caption{Experimental (Ref.~\citenum{DSLee}) and calculated atomic
coordinates for the equilibrium MnBi$_2$Te$_4$ structure.}
 \label{tab1}
\begin{center}
\begin{tabular}{cccccc}
site  & Wyckoff symbol &$x$&$y$& $z$ exp.  & $z$ calc.    \\
 \hline
Mn    & 3a      & 0 & 0 & 0.0       & 0.0      \\
Bi    & 6c      & 0 & 0 & 0.42488(4)& 0.424306 \\
Te$_1$& 6c      & 0 & 0 & 0.13333(6)& 0.134649 \\
Te$_2$& 6c      & 0 & 0 & 0.29436(6)& 0.294763 \\
 \hline
\end{tabular}
\end{center}
\end{table}

\begin{figure}[!ht]
\centering
\includegraphics[width=\columnwidth]{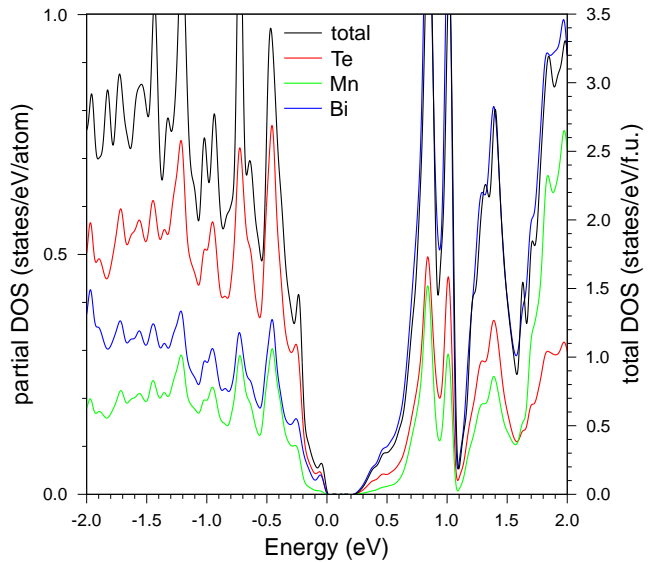}
\caption{Calculated total (black line) and partial density of states
(color lines) for the rhombohedral AFM MnBi$_2$Te$_4$. Zero energy
corresponds to the Fermi level.
 }
 \label{MBT}
\end{figure}

The calculated total density of states (DOS) for the rhombohedral
AFM MnBi$_2$Te$_4$ (Fig.~\ref{MBT}) shows that the compound is a
narrow gap semiconductor with a bandgap of 217 meV in which the
valence band maximum (VBM) as follows from the atom-projected DOS is
composed by Te and Bi $p$-orbitals while the conduction band minimum
(CBM) is formed mainly by empty Bi orbitals. The observed
experimentally for $p$-doped sample optical bandgap was estimated by
using diffuse reflectance spectrum via the Kubelka-Munk method at
room temperature to be equal to $\sim$0.4 eV \cite{DSLee}. In view
of the $p$ doping in the experiment the calculated energy gap agrees
reasonably with the measured value.

\subsection{MnBi$_2$Se$_4$}

The MnBi$_2$Se$_4$ compound was reported to be semiconducting
antiferromagnet which adopts the monoclinic crystal structure
 \cite{Ranmohotti} (see Fig.~\ref{struct} (f)). According to the X-ray Powder Diffraction
measurements the monoclinic unit cell containing four MnBi$_2$Se$_4$
formula units (Z=4) belongs to the space group $C2/m$ and is
characterized by parameters: $a$ = 13.319(3) \AA, $b$ = 4.0703(8)
\AA, $c$ = 15.179(3) \AA, and $\beta$ = 115.5(1)$^\circ$. A
triclinic Niggli-reduced cell (Z=2, see Fig.~\ref{struct} (g,h) for
this structure can be described by the parameters: $a$ =
4.07030~\AA, $b$ = 6.96350~\AA, $c$ = 13.85240~\AA, and $\alpha$ =
87.0360$^\circ$, $\beta$ = 81.5520$^\circ$, $\gamma$ =
73.0070$^\circ$.

Our calculations for this structure confirm the antiferromagnetic
configuration state for monoclinic MnBi$_2$Se$_4$. However, in
contrast to the experimental finding of Ref.~\cite{Ranmohotti} the
intrachain AFM ordering (AFM-2 in our notation, see
Fig.~\ref{struct} (i)) is the second favorable configuration after
AFM-3 (Fig.~\ref{struct} (j)) being 0.9 meV/f.u. higher in energy.
The optimized crystal cell parameters ($a$ = 4.0640~\AA, $b$ =
6.93876~\AA, $c$ = 13.81183~\AA, and $\alpha$ = 87.3886$^\circ$,
$\beta$ = 81.5400$^\circ$, $\gamma$ = 72.9718$^\circ$) as well as
atomic coordinates (see Table~\ref{tab2}) are in good agreement with
the experimental data.

\begin{table}
\caption{Experimental (recalculated for Niggli-reduced cell from the
data of Ref.~\citenum{Ranmohotti}) and calculated atomic coordinates
for monoclinic MnBi$_2$Se$_4$ structure.}
 \label{tab2}
\begin{center}
\begin{tabular}{c|cc|cc|cc}
        & \multicolumn{2}{c|}{$x$}& \multicolumn{2}{c|}{$y$}&\multicolumn{2}{c}{$z$}\\
atom    & exp.    &  calc.  &  exp.   & calc.   &  exp.   &  calc.  \\
 \hline
 Se$_1$ & 0.01145 & 0.01540 & 0.16220 & 0.15282 & 0.31490 & 0.31639 \\
 Se$_2$ & 0.98855 & 0.98460 & 0.83780 & 0.84718 & 0.68510 & 0.68361 \\
 Se$_3$ & 0.11384 & 0.11373 & 0.22719 & 0.22757 & 0.04513 & 0.04494 \\
 Se$_4$ & 0.88616 & 0.88627 & 0.77281 & 0.77243 & 0.95487 & 0.95506 \\
 Se$_5$ & 0.34158 & 0.34263 & 0.25614 & 0.25333 & 0.56070 & 0.56137 \\
 Se$_6$ & 0.65842 & 0.65737 & 0.74386 & 0.74667 & 0.43930 & 0.43863 \\
 Se$_7$ & 0.34166 & 0.33676 & 0.64172 & 0.64818 & 0.17496 & 0.17828 \\
 Se$_8$ & 0.65834 & 0.66324 & 0.35828 & 0.35182 & 0.82504 & 0.82172 \\
 Bi$_1$ & 0.28157 & 0.29149 & 0.07399 & 0.05954 & 0.86287 & 0.85749 \\
 Bi$_2$ & 0.71843 & 0.70851 & 0.92601 & 0.94046 & 0.13713 & 0.14251 \\
 Bi$_3$ & 0.35095 & 0.34844 & 0.42837 & 0.43180 & 0.36973 & 0.37133 \\
 Bi$_4$ & 0.64905 & 0.65156 & 0.57163 & 0.56820 & 0.63027 & 0.62867 \\
 Mn$_1$ & 0.00000 & 0.00000 & 0.00000 & 0.00000 & 0.50000 & 0.50000 \\
 Mn$_2$ & 0.50000 & 0.50000 & 0.50000 & 0.50000 & 0.00000 & 0.00000 \\
 \hline
\end{tabular}
\end{center}
\end{table}

However, the calculations for the MnBi$_2$Se$_4$ compound in
rhombohedral structure, performed similar to the MnBi$_2$Te$_4$ case
for FM, AFM, and NCAFM magnetic ordering, revealed that the AFM
rhombohedral structure is the lowest energy structure. It is of 39.2
meV (per formula unit) lower than the AFM-3 monoclinic structure.
Note that the cell volume in the rhombohedral structure is 1.45~\%
(per formula unit) smaller than that in the monoclinic
MnBi$_2$Se$_4$. At the same time the equilibrium rhombohedral
structure with FM and NCAFM magnetic configurations has higher
energies than the AFM ground state by only 0.6 meV and 7.8 meV,
respectively. The optimized lattice constants $a$ = 4.0782 \AA\ and
$c$ = 37.8059 \AA\ are smaller than the respective parameters of the
rhombohedral MnBi$_2$Te$_4$ owing to the smaller radius of the $Ch$
atom while the equilibrium atomic positions (Se$_1$, $z$ = 0.133816;
Se$_2$, $z$ = 0.295154; Bi, $z$ = 0.424624) are comparable with
those in MnBi$_2$Te$_4$ (see Table~\ref{tab1}).

It is worth noting that in both structures the coordination of Mn
atoms by six nearest Se atoms, which form MnSe$_6$ octahedra, is
similar. However, while in the rhombohedral structure the octahedra
are hexagonally packed, in the monoclinic phase adjacent MnSe$_6$
share edges to form one-dimensional chains along $b$ axis (see
Figs.~\ref{struct} (b) and (f)). At the same time Mn-Mn bond lengths
in hexagonal layer of rhombohedral structure and that in the chain
of monoclinic structure are very close: 4.0782 \AA\ and 4.0640 \AA,
respectively. In this regard, the fact that we found interlayer AFM
configuration for rhombohedral structure and AFM-3 one for
monoclinic structure the most energetically preferable magnetic
configurations looks reasonable. In both cases the ferromagnetic
ordering along the short Mn-Mn bonds is favorable while along the
long bonds (between hexagonal layers in rhombohedral structure and
between chains in monoclinic structure) the antiferromagnetic
coupling is preferred.

The reason for the discrepancy between the experimentally determined
and calculated crystal structure can consist in that a mixed Mn/Bi
occupancy where 6~\% of Mn occupy Bi sublattice (and vice versa)
\cite{Ranmohotti} was found in the studied sample. Such a
disordering in the Bi and Mn sublattices can presumably stabilize
the monoclinic phase in the experiment. In other words, the growth
of the rhombohedral phase can be achieved under appropriate
synthesis conditions providing suppression of the Mn/Bi intermixing.
The disordering factor can be responsible also for different type of
antiferromagnetic ordering (interchain vs. intrachain) in the
monoclinic structure.

\begin{figure}[!ht]
\centering
\includegraphics[width=\columnwidth]{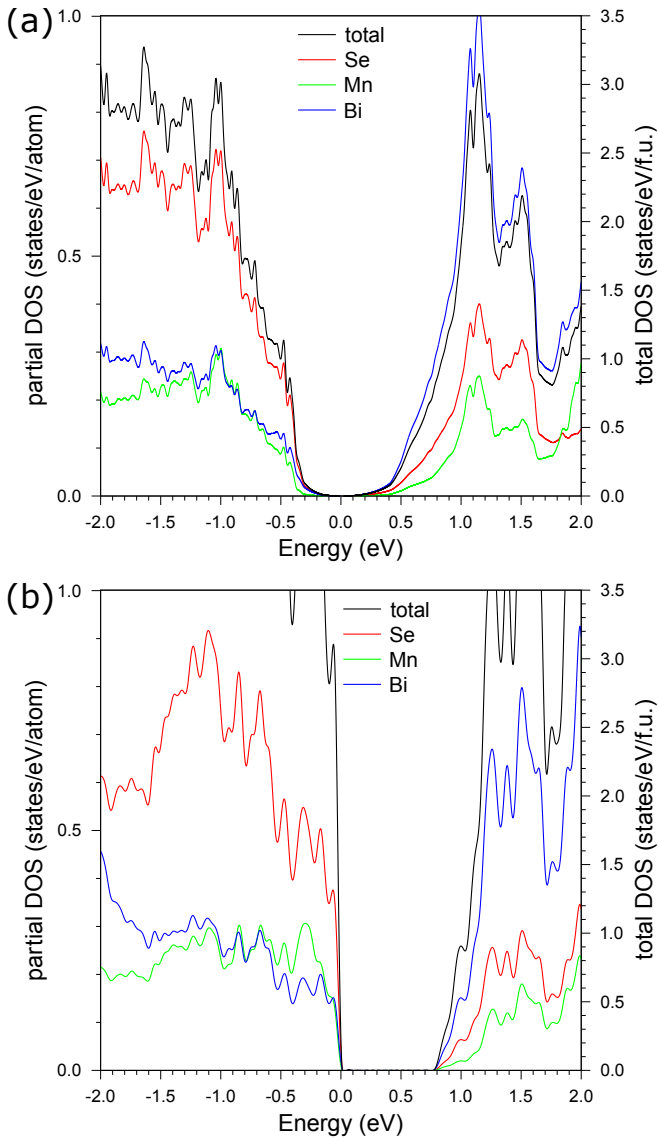}
\caption{Total and atom-projected DOS for the AFM rhombohedral (a)
and AFM-3 monoclinic (b) structures of MnBi$_2$Se$_4$.
 }
 \label{MBS}
\end{figure}

Owing to similarity in the crystal structures of rhombohedral
MnBi$_2Ch_4$ the band structure of MnBi$_2$Se$_4$ in general is
similar to that of MnBi$_2$Te$_4$. The calculated DOS for
rhombohedral AFM MnBi$_2$Se$_4$ (Fig.~\ref{MBS} (a)) demonstrates
semiconducting spectrum with very narrow bandgap of 22 meV, where,
like in the MnBi$_2$Te$_4$ case, gap edges are determined by the
$p$-orbitals of Se and Bi (VBM) and mostly Bi (CBM). The values of
the Mn magnetic moments in the rhombohedral and monoclinic phases
are also similar: 4.623 and 4.599 $\mu_{\rm B}$, respectively. Note
that, in the AFM monoclinic structure the spectrum is semiconducting
too (Fig.~\ref{MBS} (b) shows DOS for the lowest energy AFM-3 case)
in agreement with experimental finding \cite{Ranmohotti}. It has
considerably larger gap (771 meV) and Se/Bi states dominate at
VBM/CBM. Note that for other AFM configurations the DOSs are
principally the same, the gap width varies in the range of 600--771
meV and thus AFM-3 has the largest gap. At the same time the
electrical transport measurements performed on MnBi$_2$Se$_4$ single
crystals provided a band gap of 0.15 eV \cite{Nowka} that is 4-5
times smaller than the calculated value for any AFM configuration.
This discrepancy can be attributed to non-stoichiometric composition
of the measured samples.

We emphasize again that for both rhombohedral and monoclinic phases
the interlayer/interchain antiferromagnetic coupling is only
slightly more favorable than the ferromagnetic one. This result is
in line with experimental observation \cite{Ranmohotti} which
suggests the existence of residual ferromagnetic ordering in the
MnBi$_2$Se$_4$ sample. The small energy difference between magnetic
configurations is explained by the fact that in both phases the
distance between Mn layers (Mn chains) is too long for strong
magnetic exchange interaction and hence the Mn atoms of adjacent
structural blocks are magnetically coupled by indirect exchange
interactions through the Se and Bi atoms.

\subsection{MnSb$_2$Te$_4$}

The reliable data on the crystal structure of MnSb$_2$Te$_4$ are
absent with exception of the paper published in the early eighties,
Ref.~\citenum{Azhdarova}, where an unusual for the Mn$Pn_2Ch_4$
series tetragonal symmetry of the crystal structure was identified
and nothing about atomic parameters was reported. For this reason we
consider this compound within the same rhombohedral and monoclinic
phases, that are typical for related MnBi$_2Ch_4$.

The total energy calculations show that like in the MnBi$_2$Te$_4$
case the rhombohedral phase is strongly preferred. At the same time,
among magnetic configurations of the monoclinic structure, AFM-3 is
energetically favorable, as in the previously described cases of

MnBi$_2Ch_4$. The rhombohedral AFM structure is by 162.2 meV per
formula unit lower in energy than the monoclinic structure with
AFM-3 coupling. This result unambiguously indicates that
MnSb$_2$Te$_4$ can be grown in the rhombohedral phase. At the same
time, as in the other MnBi$_2Ch_4$ compounds, the energy difference
between ferromagnetic and antiferromagnetic configurations is small
for both rhombohedral and monoclinic phases and NCAFM phase is the
most unfavorable among magnetic configurations of the rhombohedral
structure. The optimized lattice constants for the rhombohedral AFM
structure are $a$ = 4.2626 \AA\ and $c$ = 39.6572 \AA, and the
atomic positions (Te$_1$, $z$ = 0.133900; Te$_2$, $z$ = 0.293205;
Sb, $z$ = 0.424323; Mn, $z$ = 0) are close to those in
MnBi$_2$Te$_4$ (see Table~\ref{tab1}).

\begin{figure}[!ht]
\centering
\includegraphics[width=\columnwidth]{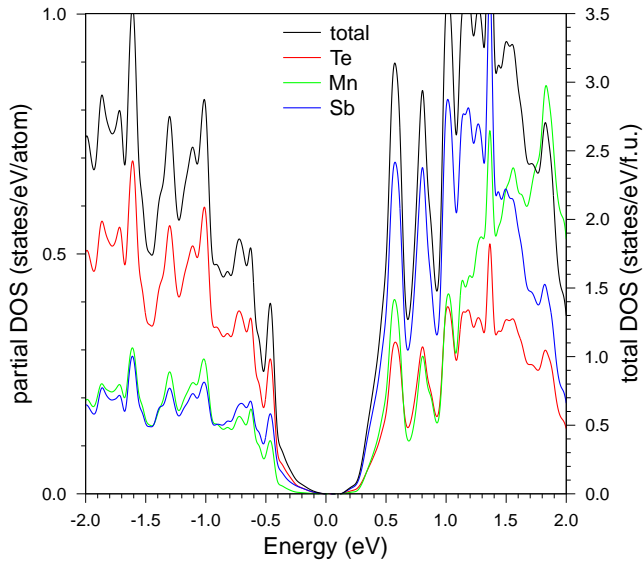}
\caption{Calculated DOS for the rhombohedral AFM MnSb$_2$Te$_4$.
 }
 \label{MST}
\end{figure}

The band structure of the MnSb$_2$Te$_4$ compound is characterized
by a bandgap of 123 meV (Fig.~\ref{MST}) that is about two times
smaller than the gap in MnBi$_2$Te$_4$. Alike the case of
MnBi$_2$Te$_4$ gap edges are contributed by the $Pn$ and Te (VBM)
and $Pn$ only (CBM) states. Owing to similarity of the crystal
structure with the MnBi$_2$Te$_4$ case and the same local atomic
surrounding for the Mn atoms the Mn magnetic moments,
4.590~$\mu_{\rm B}$, are very close to those in the Bi-containing
compound.

\subsection{MnSb$_2$Se$_4$}

The MnSb$_2$Se$_4$ compound was synthesized and its structural,
electronic and magnetic properties were studied in
Ref.~\citenum{Djieutedjeu_EJICh2011}. Single-crystal X-ray
diffraction revealed that MnSb$_2$Se$_4$ crystallizes like
MnBi$_2$Se$_4$ in the monoclinic space group C2/m with the
parameters $a$ = 13.076(3) \AA, $b$ = 3.965(1) \AA, $c$ = 15.236(3)
\AA, and $\beta$ = 115.1(3)$^\circ$ (Z=4). This structure can be
conveniently represented in the Niggli-reduced form as triclinic
cell (Z=2) with parameters $a$ = 3.96500~\AA, $b$ = 6.83200~\AA, $c$
= 13.93920~\AA, and $\alpha$ = 87.3400$^\circ$, $\beta$ =
81.8230$^\circ$, $\gamma$ = 73.1310$^\circ$.

Similarly to the above considered Mn$Pn_2Ch_4$ compounds we
performed the optimization of the crystal structure of
MnSb$_2$Se$_4$ within rhombohedral and monoclinic phases with taking
into account the magnetic ordering. Irrespective the magnetic
ordering monoclinic phase has the lower energy than rhombohedral
structure. Among the considered spin configurations for monoclinic
structure we found out that the ferromagnetic configuration is less
favorable that is in agreement with the experimental result. As it
happens for the compounds considered above the AFM-3 configuration
has the lowest energy among others. The optimized lattice parameters
for the Niggli-reduced cell are $a$ = 3.98107~\AA, $b$ =
6.87504~\AA, $c$ = 13.67025~\AA, and $\alpha$ = 87.6610$^\circ$,
$\beta$ = 81.6265$^\circ$, $\gamma$ = 73.1697$^\circ$. These
parameters as well as the atomic coordinates (see Table~\ref{tab3})
nicely reproduce the experimental structural parameters.

\begin{table}
\caption{Experimental (recalculated for Niggli-reduced cell from
data of Ref.~\citenum{Djieutedjeu_EJICh2011}) and calculated atomic
coordinates for monoclinic MnSb$_2$Se$_4$ structure.}
 \label{tab3}
\begin{center}
\begin{tabular}{c|cc|cc|cc}
        & \multicolumn{2}{c|}{$x$}& \multicolumn{2}{c|}{$y$}&\multicolumn{2}{c}{$z$}\\
atom    & exp.    &  calc.  &  exp.   & calc.   &  exp.   &  calc.  \\
 \hline
 Se$_1$ & 0.01230 & 0.01199 & 0.15530 & 0.16060 & 0.32010 & 0.31550 \\
 Se$_2$ & 0.98770 & 0.98801 & 0.84470 & 0.83940 & 0.67990 & 0.68450 \\
 Se$_3$ & 0.11430 & 0.11647 & 0.22890 & 0.22199 & 0.04250 & 0.04503 \\
 Se$_4$ & 0.88570 & 0.88353 & 0.77110 & 0.77801 & 0.95750 & 0.95497 \\
 Se$_5$ & 0.34510 & 0.34087 & 0.25650 & 0.25946 & 0.55330 & 0.55880 \\
 Se$_6$ & 0.65490 & 0.65913 & 0.74350 & 0.74054 & 0.44670 & 0.44120 \\
 Se$_7$ & 0.34460 & 0.33405 & 0.63850 & 0.65225 & 0.17230 & 0.17967 \\
 Se$_8$ & 0.65540 & 0.66595 & 0.36150 & 0.34775 & 0.82770 & 0.82033 \\
 Sb$_1$ & 0.27380 & 0.28382 & 0.08580 & 0.07050 & 0.86660 & 0.86184 \\
 Sb$_2$ & 0.72620 & 0.71618 & 0.91420 & 0.92950 & 0.13340 & 0.13816 \\
 Sb$_3$ & 0.35540 & 0.35164 & 0.41520 & 0.42240 & 0.37400 & 0.37437 \\
 Sb$_4$ & 0.64460 & 0.64836 & 0.58480 & 0.57760 & 0.62600 & 0.62563 \\
 Mn$_1$ & 0.00000 & 0.00000 & 0.00000 & 0.00000 & 0.50000 & 0.50000 \\
 Mn$_2$ & 0.50000 & 0.50000 & 0.50000 & 0.50000 & 0.00000 & 0.00000 \\
 \hline
\end{tabular}
\end{center}
\end{table}

\begin{figure}[!hb]
\centering
\includegraphics[width=\columnwidth]{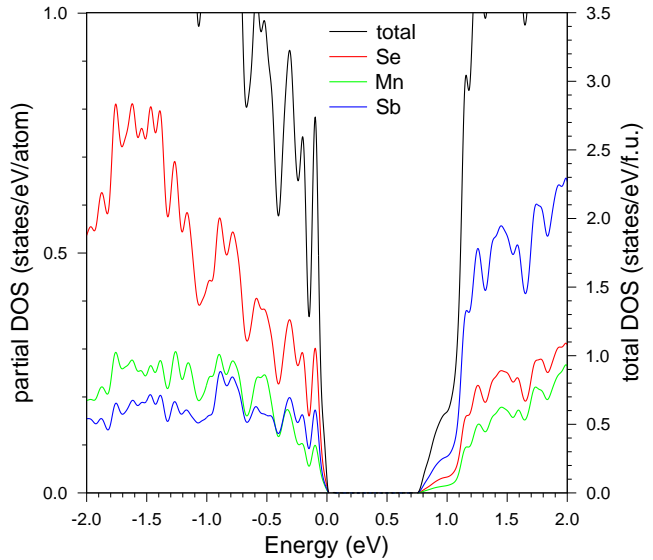}
\caption{Calculated DOS for the monoclinic structure of
MnSb$_2$Se$_4$ with AFM-3 magnetic ordering.
 }
 \label{MSS}
\end{figure}

Like in case of other Mn$Pn_2Ch_4$ the antiferromagnetic ordering of
the Mn magnetic moments is only slightly more favorable than the
ferromagnetic configuration. According to the magnetic
susceptibility measurements performed in
Ref.~\citenum{Djieutedjeu_EJICh2011} the interaction between the Mn
magnetic moments is predominantly antiferromagnetic, however, the
slight increase in the susceptibility observed below 5 K was
explained by the existence of a residual ferromagnetic ordering in
the sample at very low temperatures. Our results showing small
energy difference between ferro- and antiferromagnetic
configurations confirm the competition between AFM and FM ordering
in the compound. The calculated magnetic moment on the Mn atoms for
the most favorable AFM-3 structure is 4.596~$\mu_{\rm B}$, which is
almost the same as that in monoclinic MnBi$_2$Se$_4$.

Experimental estimation with using the diffuse reflectance infrared
spectroscopy measurements at room temperature gave the value of the
bandgap in the MnSb$_2$Se$_4$ sample of $\sim$0.32 eV
\cite{Djieutedjeu_EJICh2011}. On the other hand, from the
temperature dependent electrical resistivity measurements the
bandgap value was estimated to be of 0.52 eV
\cite{Djieutedjeu_EJICh2011}. Our calculations of the electronic
structure provide a bandgap, which is closer to the experimental
value derived from the charge-transport data. As can be seen from
the calculated DOS for monoclinic AFM-3 structure presented in
Fig.~\ref{MSS}, in MnSb$_2$Se$_4$ the gap is of 757 meV that is
comparable with the value in the monoclinic MnBi$_2$Se$_4$ (772
meV). Like in the case of MnBi$_2$Se$_4$ Se/Bi states dominate at
the gap edges.

\section{Discussion and concluding remarks}

In summary, we have performed DFT calculations of electronic,
magnetic and crystal structure of the Mn$Pn_2Ch_4$ series of ternary
transition metal chalcogenides ($Pn$ = Sb, Bi; $Ch$ = Se, Te). All
Mn$Pn_2Ch_4$ compounds were considered within the rhombohedral and
monoclinic phases, which were shown experimentally to be typical for
some compounds of the series. The FM, interlayer AFM and NCAFM spin
structures have been taken into account in the calculations for
rhombohedral structure and FM and four different AFM configurations
for monoclinic phase. The obtained total energies for Mn$Pn_2Ch_4$
phases are summarized in the Table~\ref{tabEnrg}. We have found that
the compounds containing the heavier chalcogen atom, Te, show a
strong trend to adopt the layered rhombohedral structure.

\begin{table}[!hb]
\caption{Relative total energies (in meV) per formula unit (zero
energy corresponds to rhombohedral AFM case) for different magnetic
states of rhombohedral and monoclinic structures of Mn$Pn_2Ch_4$
compounds.}
 \label{tabEnrg}
\begin{center}
\begin{tabular}{c|ccc|ccccc}
               & \multicolumn{3}{c|}{rhombohedral} & \multicolumn{5}{c}{monoclinic} \\
compound       & AFM & FM  & NCAFM  & FM    & AFM-1  & AFM-2  & AFM-3  & AFM-4   \\
 \hline
MnBi$_2$Te$_4$ & 0.0 & +4.5& +11.8  & +206.4& +205.6 & +204.6 & +202.7 & +205.2 \\
MnBi$_2$Se$_4$ & 0.0 & +0.6& +7.8   & +43.0 & +41.7  & +40.1  & +39.2  & +40.6  \\
MnSb$_2$Te$_4$ & 0.0 & +1.3& +11.2  & +165.9& +165.1 & +164.7 & +162.2 & +165.3 \\
MnSb$_2$Se$_4$ & 0.0 & +0.8& +8.1   & -11.3 & -12.7  & -13.7  & -15.1  & -13.3  \\
 \hline
\end{tabular}
\end{center}
\end{table}


Thus, our results confirm the structure of MnBi$_2$Te$_4$, recently
determined by the experiment and predict the similar crystal
structure for MnSb$_2$Te$_4$, which has not been studied in details
earlier. For the compound with lighter both pnictogen and chalcogen
atoms, MnSb$_2$Se$_4$, our result, predicting the lowest energy
state for monoclinic structure, is in agreement with experimental
data. A similar change in the crystal structure from the
rhombohedral layered phase to the monoclinic structure was reported
earlier for compounds with substitution between Se and Te in
FeSb$_2$Te$_{4-x}$Se$_x$ ($x$ = 1,2,3,4) \cite{DjieutedjeuPhD}. At
the same time our total energy calculations for MnBi$_2$Se$_4$,
which contains heavier pnictogen and lighter chalcogen atoms,
predicting the rhombohedral phase as the stable structure contradict
the experimentally determined monoclinic structure. However, it
should be noted that the energy gain for the rhombohedral over
monoclinic structure in MnBi$_2$Se$_4$ is five times smaller than in
the MnBi$_2$Te$_4$ compound (39.2 vs. 202.7 meV, see
Table~\ref{tabEnrg}). This discrepancy can be an indication that the
rhombohedral phase of MnBi$_2$Se$_4$ can be obtained at the growth
conditions that are different from those used in
Ref.~\cite{Ranmohotti}. E.g. it could probably be stabilized by
using MBE. We have also shown, that irrespective of the crystal
structure, all compounds of the Mn$Pn_2Ch_4$ family are
antiferromagnetic semiconductors. In accordance with the
experimental findings for Mn$Pn_2$Se$_4$, indicating the competition
between AFM and FM ordering, the energy gain for the AFM coupling
with respect to ferromagnetic state is very weak -- it is just a few
meV per formula unit or even smaller. The small energy difference
between antiferromagnetic and ferromagnetic configurations is
explained by the fact that in both phases the distance between Mn
layers is too long for strong magnetic exchange interactions.  At
the same time among considered antiferromagnetic configurations for
monoclinic structure AFM-3, in which ferromagnetically ordered Mn
chains couple to each other, is the most favorable. Owing to
similarity in the local atomic surrounding for the Mn atoms the Mn
magnetic moments are almost the same through the Mn$Pn_2Ch_4$ series
regardless of the structure. On the other hand, the type of crystal
structure, rhombohedral or monoclinic, significantly influences the
bandgap width. In the rhombohedral Mn$Pn_2Ch_4$ phases the gap is of
22-217 meV while in the compounds with monoclinic structure this
value is significantly larger and amounts to about 750-770 meV.


%

\section*{References}


%

\end{document}